\documentclass[authoryear]{elsarticle}
\usepackage{epsfig,graphicx,aas_macros}
\usepackage{graphics,graphicx}
\usepackage{amsmath}
\usepackage[font=small,labelfont=bf]{caption}
\usepackage[authoryear]{natbib}
\begin{document}

\begin{frontmatter}

\title{Post-flare Formation of the Accretion Stream and a Dip in Pulse Profiles of LMC~X--4}
\author[mymainaddress,mysecondaryaddress]{Aru Beri \corref{mycorrespondingauthor}}
\cortext[mycorrespondingauthor]{Corresponding author}
\ead{a.beri@soton.ac.uk}
\author[mysecondaryaddress]{Biswajit Paul}
\address[mymainaddress]{Royal~Society--SERB Newton International Fellow,~School of Physics and Astronomy,~University of Southampton,~Southampton,~Hampshire SO17~1BJ,~UK}
\address[mysecondaryaddress]{Raman Research Institute, C. V. Raman Avenue, Sadashivanagar, Bangalore-560012, India }

\begin{abstract}
 We report here a pulse profile evolution study of an accreting X-ray pulsar LMC~X--4
during and after the large X-ray flares
using data from the two observatories \emph{XMM-Newton} and \emph{RXTE}.
During the flares, the pulse profiles were found to have a significant phase offset
in the range of 0.2-0.5 compared to the pulse profiles immediately before or after the flare.
Investigating the pulse profiles for about $10^5$~seconds after the flares, it was found 
that it takes about 2000-4000~seconds for the modified accretion column
to return to its normal structure and formation of an accretion stream
that causes a dip in the pulse profile of LMC~X--4.
 We have also carried out pulse phase resolved spectroscopy of LMC~X--4 in narrow phase bins using data 
 from \textsc{EPIC}-pn and spectroscopically confirmed the pulsating nature
 of the soft spectral component, having a  pulse fraction and phase different from that of the power-law component.
 
\end{abstract}


\end{frontmatter}
LMC~X--4 is a persistent,~disk-fed, X-ray binary pulsar which consists of a 1.25 $M_{\odot}$
neutron star and a 14th magnitude O-type star of mass $\sim$15$M_{\odot}$ 
\citep{Kelley83,vander07} in the Large Magellanic Cloud.
This high-mass system exhibits pulse period and an orbital period of $\sim$~13.5~s,
and $\sim$ 1.4~days respectively \citep{Kelley83}.
A periodic superorbital intensity modulation at $\sim$~30.5~day
during which the X-ray intensity varies by the
factor of $\sim$~60 between high and low states has also been observed in this source \citep{Lang81, Paul02a}. \\
\par
LMC~X--4 is one of the few X-ray binaries that shows large X-ray flares \citep{Levine91, Levine00,
Moon01, Moon03}.~X-ray flares in LMC~X--4 are believed to occur due to increased mass accretion because of 
Rayleigh-Taylor instability 
of plasma in the accretion disk \citep{Moon03}.
Sometimes the flare X-ray luminosity reaches 
up to $\sim$$10^{39}$ergs/sec \citep{Levine00, Moon03}.
This super-Eddington emission is understood to be due to
inhomogeneous accretion columns during the flares \citep{Moon03}.
During the LMC~X--4 flares,~pulse profiles are broad and sinusoidal in shape
\citep{Levine91,Levine00,Moon03}.~\citet{Moon03} suggested that during flares most of the flare emission
is transported via a fan beam in the direction perpendicular
to the magnetic field.
However,~this is in contrast to the pulse profiles observed~outside the X-ray flares.
Pulse profiles of LMC~X--4 in a soft energy band show complex 
structures with dips in it \citep{Levine91, Levine00, Paul02, Paul04}.~Significant changes
in the pulse profiles of LMC~X--4, during large flares suggest
 that it is a rare source to allow
study of evolution of pulse profile during the transition between flaring and non-flaring states
and to understand accretion flow in the system.
During X-ray flares,
the accretion column and the beaming of X-ray emission could	
be altered. \\
\par
Dips are the sharp drops in intensity over a narrow phase range of ${\Delta}{\phi}~{\approx}~0.1$
\citep[see e.g.,][]{Greenhill98, Giles00, Galloway01,
Devasia10, Devasia11a, Devasia11b, Usui12, Naik13}.
The depth of a dip can attain a very high value, 80-100$\%$ of the pulse maximum \citep{Giles00}.
These are often attributed to the eclipse of the emitted radiation
by the optically thick material in the accretion column \citep{Cemeljic98}.
Certain viewing angle configurations
of the systems can allow the passage of the accretion column through our line of sight at some pulse phases, giving rise
to reprocessing of the emitted radiation from the column, where the softer low energy photons
can either get absorbed or scattered out of our line of sight. \\
\par
Dips in the pulse profiles are often 
predominant at energies below 15~keV \citep{Devasia11b, Maitra12, Maitra_13, Devasia14}.
However, there exist a few exceptions 
such as GX~1+4 \citep{Giles00},~EXO~2030+375 \citep{Naik13a} and GS~1843--02 \citep{Devasia14}
in which dips are observed even at energies above 15~keV.
~Pulse profiles of a particular source may have more than one such features and these
multiple dips are understood as the indication of more than one accretion stream \citep{Devasia14}. 
They may also provide a unique opportunity to investigate the
timescale required for the formation of the accretion stream that causes a
dip in its pulse profiles after the accretion region and the beaming etc. is disturbed during the flares. \\ 
\par
The X-ray spectrum of LMC~X--4 in 0.1--100~keV band is described by a power law with a high-energy cutoff,~a
soft X--ray excess, and an iron emission line \citep{La01, Naik03}.
The soft X-ray excess is believed to originate from the reprocessing of hard X-rays
from the neutron star by the inner accretion disk and is detectable only in those sources which have
low value of hydrogen column density like SMC~X--1, HerX--1, LMC X-4 \citep{Paul02,Hickox04}.
Pulse Phase Resolved Spectroscopy performed using data from the \emph{ROSAT}, \emph{Ginga}, \emph{ASCA} and \emph{BeppoSAX} observatory
revealed some evidence that the
soft spectral component of LMC~X--4 is pulsating with a different pulse phase with respect to the power law component \citep{Woo96, Paul02, Paul04}.
However, this can be better investigated using data from the \textsc{EPIC}-pn instrument aboard \emph{XMM-Newton}. \\
\par
In this paper, we report a very detailed pulse profile evolution study of LMC~X--4,
using data from two observatories,~\emph{XMM-Newton} and \emph{RXTE}. The results
obtained from pulse phase resolved spectroscopy of LMC~X--4 using data from \textsc{EPIC}-pn are
also presented.
The paper is organized as follows~: 
  Section-2 gives the details of the observations used in this work, section-3 
  describes pulse profile studies performed during flaring and non-flaring states
  using data from \textsc{EPIC}-pn and \emph{RXTE}-PCA.
  We describe spectral studies of this source in section-4, where we investigate the behavior of 
  soft excess in narrow phase bins.
  The last section discusses the results obtained from the analysis.

\section{Observations and Data Reduction}

We have selected 
those observations which include several large flares and a large continuous stretch of persistent emission
after the flares.~(We refer to the persistent emission in this paper
as a state which has no flares and eclipse.)  \\
\par
We have used data from the \textsc{EPIC}-pn onboard \emph{XMM-Newton} and the \textsc{PCA} of \emph{RXTE} for this work.
Table-1 shows the log of observations used in this work.~Here, we would like to mention that 
 these observations have been used before, but have never been analyzed for the purpose of pulse profile 
 evolution study.
 \citet{Neilsen09} used the same \emph{XMM-Newton} observation for performing spectroscopic studies while the
 other two \textsc{PCA} observations were used for the detailed examination of flares \citep[see][]{Moon03}. 
 Very recently, the same observations were used by \citet{Molkov17}
for the spin period measurements. \\
\par
The European Photon Imaging Camera (\textsc{EPIC}), and the Reflection Grating Spectrometer (\textsc{RGS}) are two X-ray
 instruments aboard \emph{XMM-Newton} which operate in the energy range of 0.1-15~keV. 
The \textsc{EPIC} consists of two MOS \citep{Turner01} and one
pn \citep{Struder01} CCD arrays having moderate spectral resolution and 
 a time resolution in the range of micro-seconds to 2.6 seconds depending on the instrument and the mode of observation. 
An optical Monitor~(\textsc{OM}) performs simultaneous optical/UV observations. \\
\textsc{EPIC}-pn data were collected in the small window mode with thick filter.
Frame time of these observations was 6~ms.~This observation belongs to the high state of the $\sim$~30.5~days superorbital cycle \citep{Neilsen09}. \\ 
 
For the reduction and extraction purposes, we have used
\emph{HEASOFT~6.12} and \emph{SAS-12.0.1}.~Standard filters were applied to the extraction of the
cleaned \textsc{EPIC}-pn events.
To check whether the data was affected with soft proton flaring, 
a light curve was extracted by selecting events
with PATTERN=0 and energy in the range of 10-12~keV.
Thereafter, a good time interval~(gti) with rate~$\leq$0.4 was created.~This gti was then used to obtain the 
filtered events.~The \emph{SAS} tool \textit{evselect} was used to perform the particle background check
and for filtering out background flares prior to analysis.
 A circular region of 40~arcsecond radius was selected around the
source centroid for the extraction of source events.
The \emph{SAS} tool \textit{epatplot} was used to detect presence of any 
possible pile-up in the data.
For the persistent emission, the count rate was slightly higher than the
maximum count rate limit for \textsc{EPIC}-pn with a small window mode~(25~counts/s),
 while for the case of flaring events, data were heavily piled up.
Therefore, we have performed a pile-up correction by removing a radius of 7.5~arcseconds from the core of the PSF.
This annular source region file was then used for light curve extraction during the flares.
However, we have used all the pixels of source events for extraction of light curves during the persistent emission.
 We have used the annular source region file for extraction of source spectra
during the persistent emission.
PATTERN $\leq$ 4 and FLAG=0 were used as selection criteria.
Two off-source regions were used with a radius of 40~arcseconds
for the background spectra and light curve extraction.
An updated version of the current calibration files 
\footnote[1]{http://xmm2.esac.esa.int/external/xmm\texttt{\_sw\_cal}/calib/ccf.shtml} 
was used for reprocessing of the data and creation of the response files. \\

The \textsc{PCA} instrument on board \emph{RXTE}
consists of five Proportional Counter Units
(PCUs) covering an energy range of 2--60 keV with an effective area of 6500 cm$^{2}$ \citep{Jahoda96, Jahoda06}.
Both the \emph{RXTE} observations (given in Table-1) have 6 pointings.
For screening of data, the following filtering criteria were applied: time since South Atlantic Anomaly (SAA) 
was greater than 10~minutes, pointing offset was less than 0.02~degrees, and the
earth elevation angle was greater than 10~degrees. Only the data obtained with two or more active
Proportional Counter Units (PCUs) were used.
\textsc{PCA} data collected in Good Xenon1 and Good Xenon2 modes were used to generate the source light curves.
The  tool \textsc{runpcabackest} was used to estimate the background, assuming a source model
as suggested by \emph{RXTE} GOF\footnote{(http://heasarc.gsfc.nasa.gov/docs/xte/pca$_{-}$news.html)}
and subsequently background light curves were generated. \\
 
The arrival times of photons from both \textsc{EPIC}-pn and \textsc{PCA} were first converted to the solar system barycenter.
Due to the short orbital period of LMC~X--4, pulses may lose coherence within a relatively
short timescale of a few thousand seconds. Therefore, the arrival time of each photon was corrected
for the binary motion considering the semiamplitude to be 26.3~seconds and the mideclipse time 
as per its orbital evolution rate \citep{Paul04}.~The orbital parameters used for this correction
are given in Table-\ref{orbit}.

\begin{table}
\centering
\caption{Log of observations}
\begin{tabular}{@{}lccc@{}}
\hline
\hline
Observatory             &  Year & Observation ID &   Total Exposure (ks)  \\
\hline
\emph{\emph{XMM-Newton}}  &      &                &                     \\
& 2003     &  0142800101     & 113  \\
\hline
\emph{RXTE}             &          &              &          \\
             & 1996     &  P10135         & 170   \\
             & 1999     &  P40064         & 150   \\

\hline
\end{tabular}
\label{OBS}
\end{table}

\begin{table}
\centering
\caption{Orbital Parameters of LMC~X--4 \citep{Paul04} used for the correction due to its orbital motion.}
\begin{tabular}{@{}lccc@{}}
\hline
\hline
Parameter                   &  Units      & Value   \\
\hline
$a_{x}sini$                 & lt-secs     &  $26.33\pm0.02$      \\
$P_{orb}$                   &  days       &  $1.40839\pm0.00002$               \\
$\dot{P_{orb}}/P_{orb}$     & $yr^{-1}$   &  $(9.69\pm0.07)$$\times10^{-7}$           \\
$T_{mideclipse}$ (pn)       &  MJD        &  52892.4844(17) \\
$T_{mideclipse}$ (P10135)   &  MJD        &  50315.1225(14) \\
$T_{mideclipse}$ (P40064)   &  MJD        &  51531.9727(25) \\

\hline
\end{tabular}
\label{orbit}
\end{table}

\section{Timing Analysis with \textsc{EPIC}-pn and \textsc{PCA}}

\begin{figure*}
\centering
\includegraphics[height=6.5in,width=30.0cm,keepaspectratio]{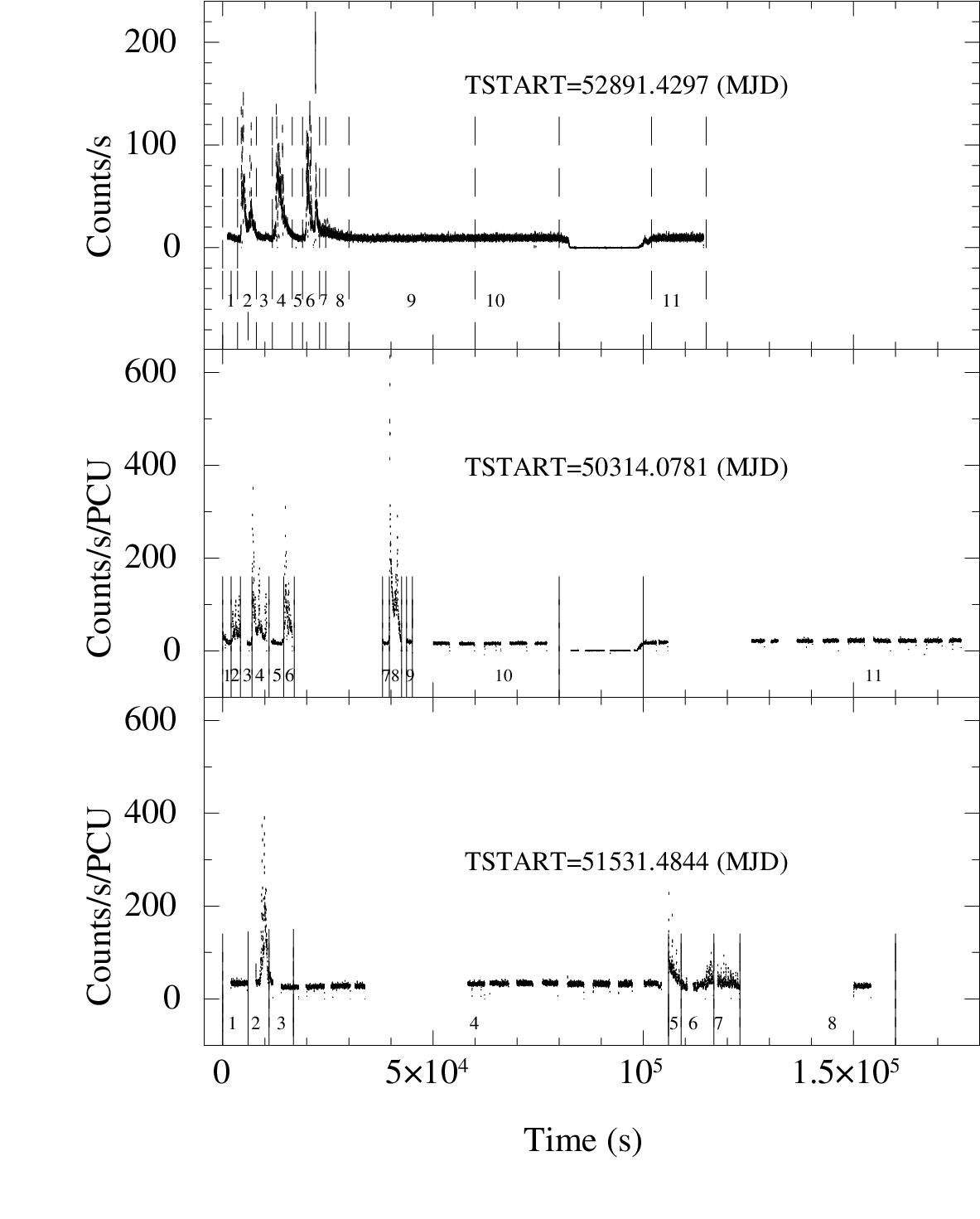}
\caption{The top panel shows the light curve of LMC~X--4, 
obtained using data from \textsc{EPIC}-pn after removing piled-up pixels, the middle panel 
shows the light curves created using the observation made with \emph{RXTE}-\textsc{PCA}
with ID-P10135 while the bottom panel shows the light curves obtained using \textsc{PCA}
data with observation ID-40064.~The 2-10~keV band was used for creating
these light curves.
The plotted light curves have been binned with a bin size equal to the pulse period
of LMC~X--4,$\sim$13.5~s.
}
\label{LC}
\end{figure*}

\subsection{Light Curves}

The time series obtained from the observation made with the \textsc{EPIC}-pn has flares,~persistent
emission and an eclipse~(top panel of Figure-\ref{LC}).~The flares lasted for nearly 20~ks.
The light curve extracted in the 2-10~keV band, excluding the piled up pixels showed an
average count-rate of $\sim$~9~counts/s during the persistent emission and a peak count rate during the flares
up to 20 times the mean count rate during the persistent emission. An exponential decay timescale for each of the flares~($2$,~$4$,~$6$) (Figure~\ref{LC})
 is 470~s, 880~s and 700~s respectively. \\

Light curves in the 2-10~keV band were created using the \textsc{PCA} data of the observations mentioned
in the Table-\ref{OBS}.
Light curves created using the observation with ID~P10135 showed four flares,
persistent emission and an eclipse (see bottom panel of Figure-\ref{LC}).
The flares lasted for about 40~ks. The average count rates per PCU observed during the persistent
emission is 30 counts/s and the flares reached a peak count rate of about 650 counts/s/PCU.
An exponential decay timescale measured for the four flares~($2$, $4$, $6$ \& $8$) seen in observation~P10135
 are about 160~seconds, 140~seconds, 540~seconds and 300~seconds respectively. \\
 
Another \textsc{PCA} observation with ID--P40064 also showed four flares
in addition to the persistent emission.
These four flares were, however, not continuous but had a time gap between the
first and the other three (see middle panel of Figure-\ref{LC}).
The average count rate per PCU observed during
persistent emission is 26 counts/s in 2-10~keV band.
Flares reached the maximum of 400~counts/s/PCU for a light curve with binsize 13.5~seconds.
In the observation~P40064, an exponential
 decay timescales measured for the two high intensity flares~($2$ and $5$) are 480~seconds and 500~seconds, respectively. \\

\subsection{Energy-Resolved Profiles}

We have used the light curves during persistent emission 
for the spin period measurement. 
 For the \textsc{EPIC}-pn observation, persistent emission commenced about 5000~seconds
 after the end of last flare~($segment~6$).~For the \textsc{PCA} observation with ID~P10135,
 we have used data 2000~seconds after the end of last flare~(marked as $8$ in the second panel of Figure-\ref{LC}).
 We have included all segments of the persistent emission excluding the intermediate
 flares~(segments:~$2$,~$5$,~$6$,~$7$) for the other \textsc{PCA} observation with ID~P40064. \\
 \par
The spin period was measured to be 13.49624(7),~13.50912(1)
and~13.49642(2)~seconds using the pulse folding $\chi^2$ maximization technique 
applied to the light curves during persistent emission from the \textsc{EPIC-pn} and the
two \textsc{PCA} observations, P10135 and P40064 respectively.
1~$\sigma$ uncertainties in the pulse periods were estimated opting the same technique as used by \citet{Naik05}.
We notice that spin-periods measured are consistent 
with those reported by \citet{Molkov17}.
The respective period of each observation 
was used to create energy resolved pulse profiles using both the flaring and persistent emission
data, using the \textsc{FTOOL} task {\it{efold}}.  \\
\par
\begin{figure*}
\centering
\begin{minipage}{0.45\textwidth}
\includegraphics[height=6.5in,width=7.cm,keepaspectratio]{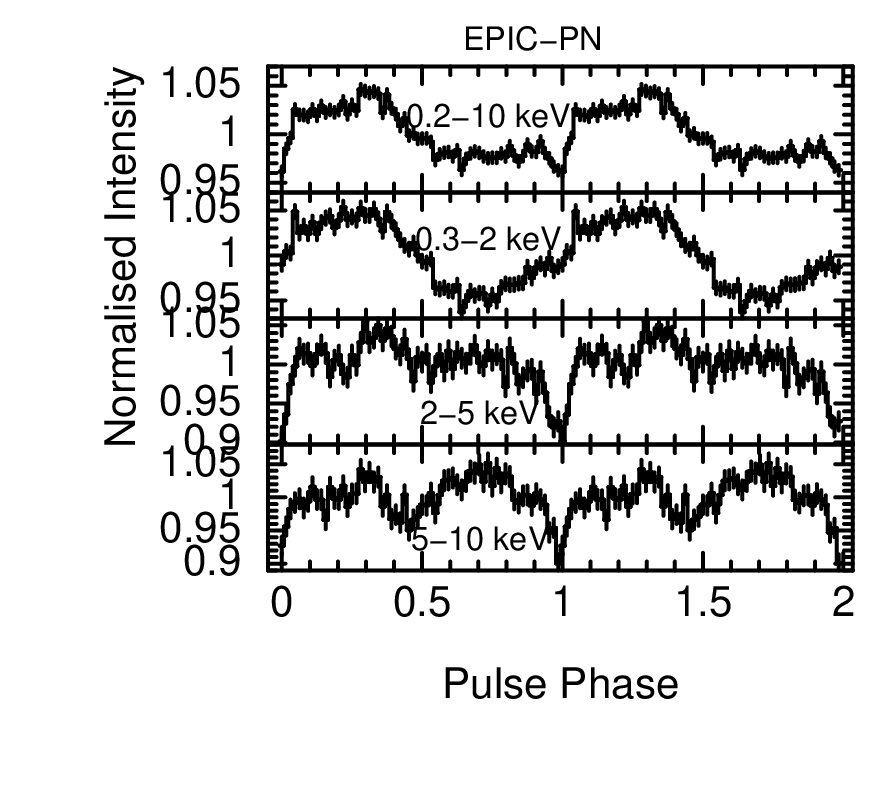}
\end{minipage}
\hspace{0.05\linewidth}
\begin{minipage}{0.45\textwidth}
\includegraphics[height=6.5in,width=7.cm,keepaspectratio]{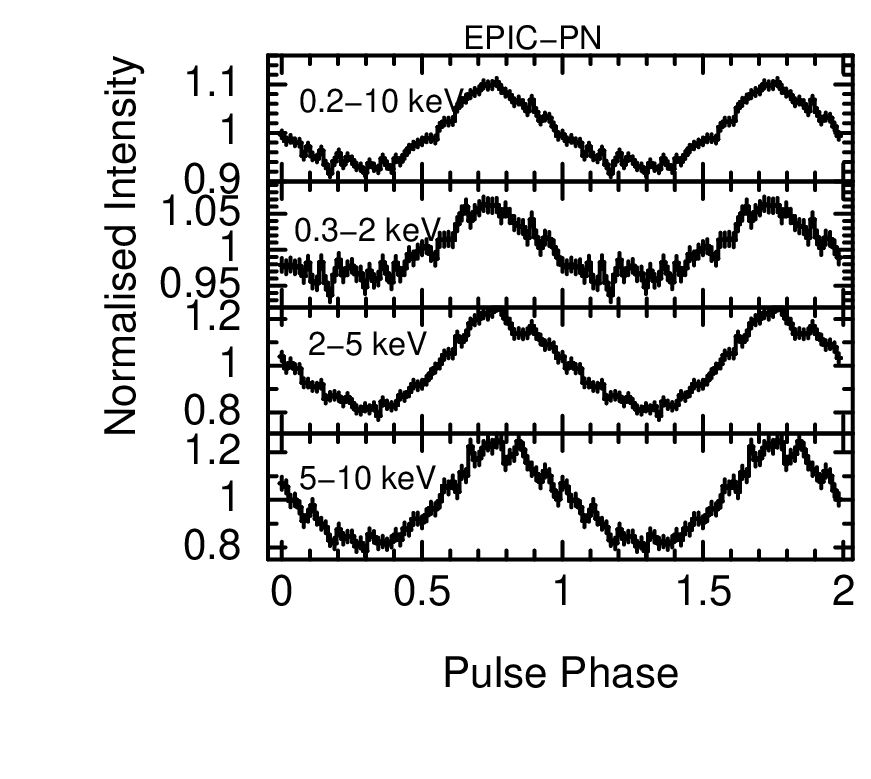}
\end{minipage}
\caption{Left:~Energy resolved pulse profiles created using the \textsc{EPIC}-pn data during the persistent state,~binned into 64 phasebins.~Right:~Energy resolved pulse profiles created using data during the flares of same observation.}
\label{pp}
\end{figure*}

Phase zero for each observation was determined with respect to the 
sharp, deep dip observed in the pulse profiles during persistent emission (deep dip which settled in phase after large flares).
In Figure-\ref{pp},~we show representative energy resolved pulse profiles, created using the \textsc{EPIC}-pn 
data.
The phases of the pulse profile for all the three observations are adjusted 
with the dip appearing at phase zero.
In the 0.3-2~keV band,~the pulse profile during persistent emission is single peaked with a shoulder
type structure.
The dip feature is seen only at energies above 2~keV. 
These pulse profile shapes are quite similar to the previously reported profiles
created using observations made with \emph{Ginga}, \emph{RXTE}, \emph{ASCA}, \emph{BeppoSAX} and \emph{Suzaku} \citep{Levine91,Levine00,
Paul02, Paul04, Hung10}.~Similar dips were also observed in the pulse profiles during the persistent emission, created using data from the other two \textsc{PCA} observations.
Pulse profiles during the flares, on the other hand, 
exhibit simple sinusoidal shapes in all the energy bands~(Figure~\ref{pp}). This is consistent with the known facts of energy resolved pulse 
profiles during flares of LMC~X--4 \citep{Levine91,Levine00,Moon03}. \\

\subsection{Evolution of dips in the pulse profiles}

\begin{figure*}
\centering
\begin{minipage}{0.45\textwidth}
\includegraphics[height=6.5in,width=7.cm,keepaspectratio]{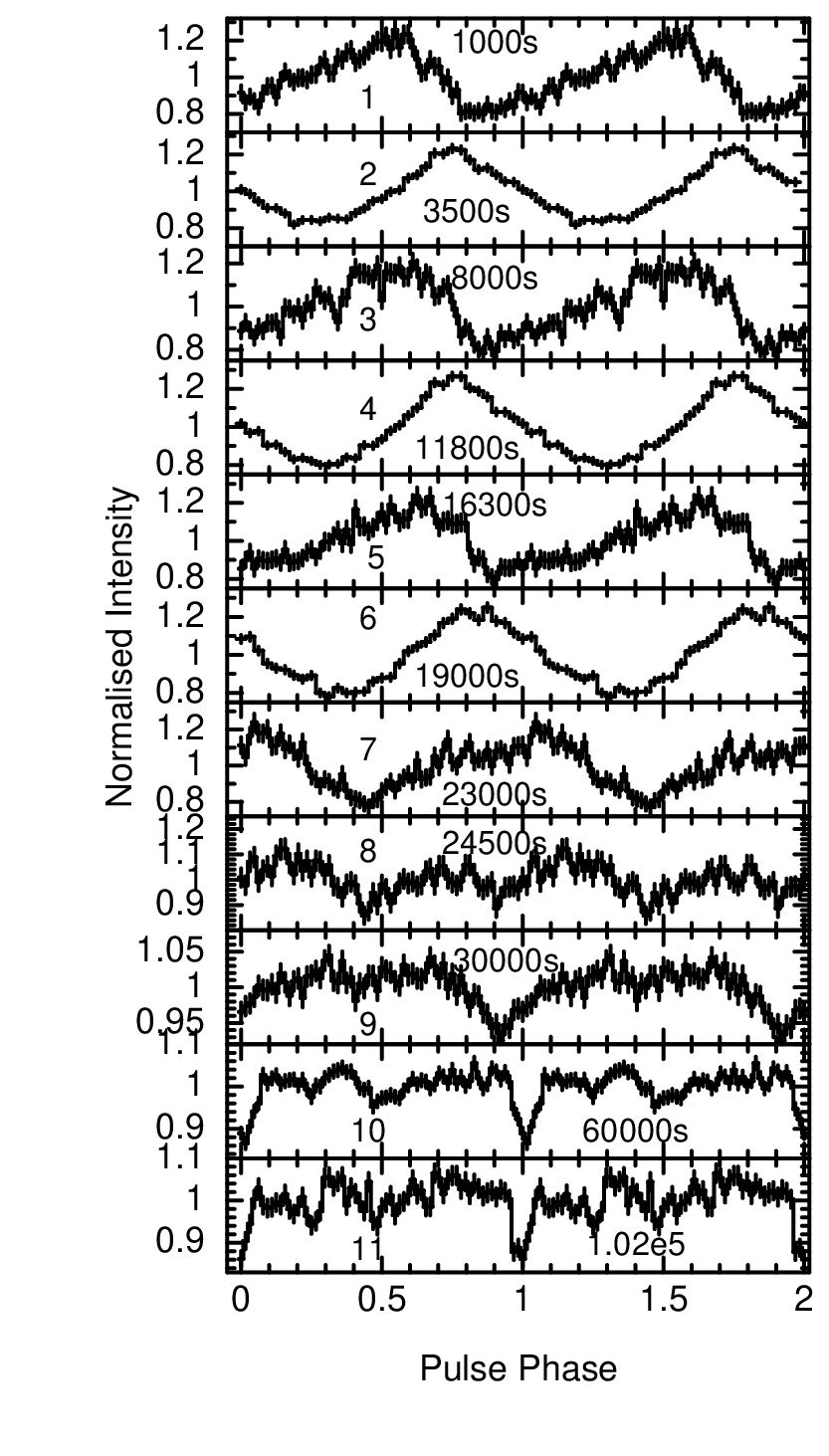}
\end{minipage}
\hspace{0.05\linewidth}
\begin{minipage}{0.45\textwidth}
\includegraphics[height=6.5in,width=7.cm,keepaspectratio]{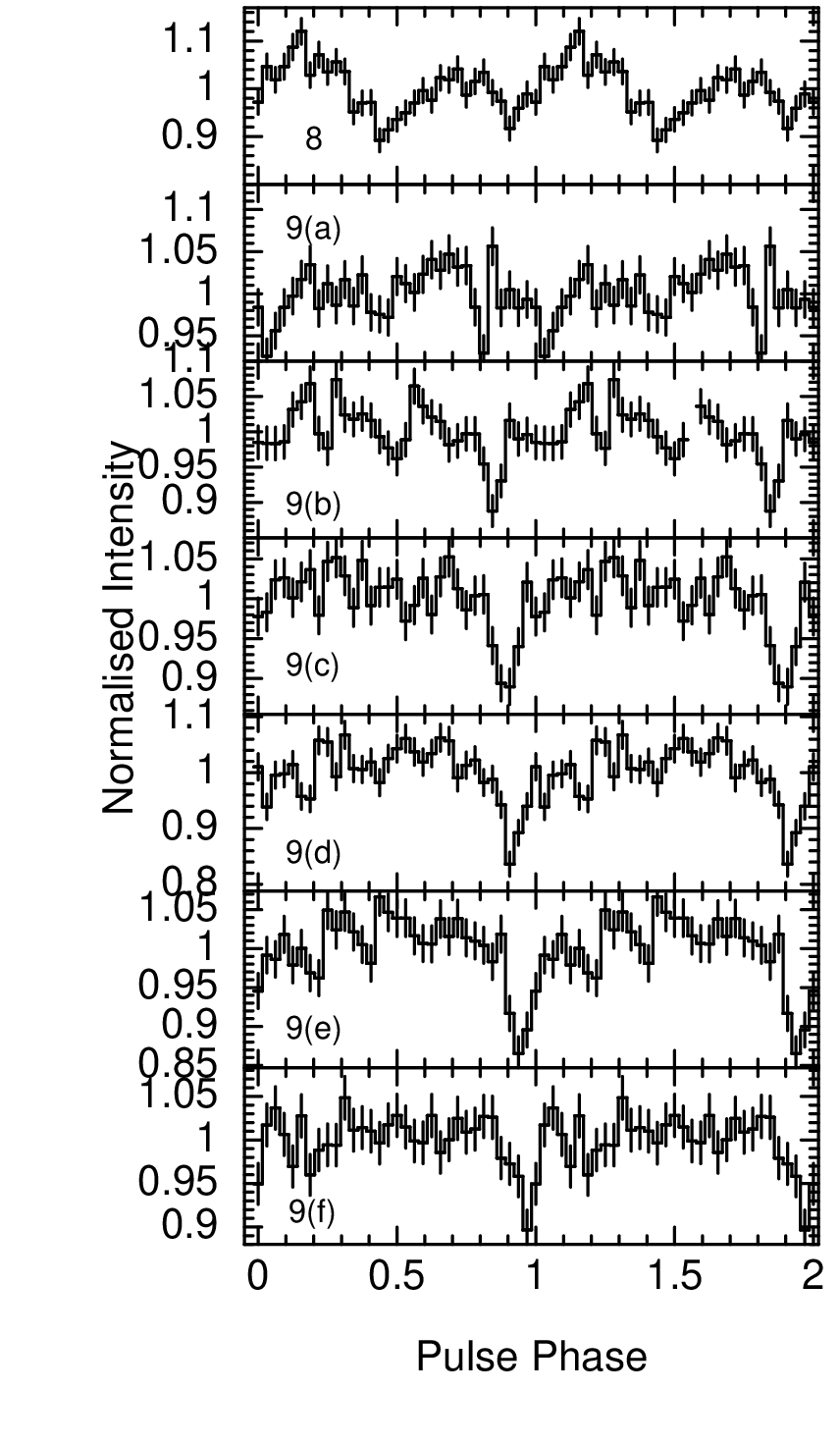}
\end{minipage}
\caption{The left-hand plot shows the pulse profiles created with \textsc{EPIC}-pn data 
in short segments as marked in the top panel plot of Figure-\ref{LC}.
Each panel is marked with the time offset relative to the start of the observation.
The profiles are created using time series in the 2-10~keV band and binned into 64 phasebins while the plot on the right 
shows the pulse profiles created using the sub-segments of segment $9$, binned into 32 phasebins,
in order to demonstrate the time required for the evolution of 
dips after the flares.~Segment~9 was divided into 6 short segments, each with 5~ks of exposure.}
\label{pp-evol}
\end{figure*}

 \begin{figure*}
 \centering
 \includegraphics[height=5.5in,width=8.5cm,angle=0,keepaspectratio]{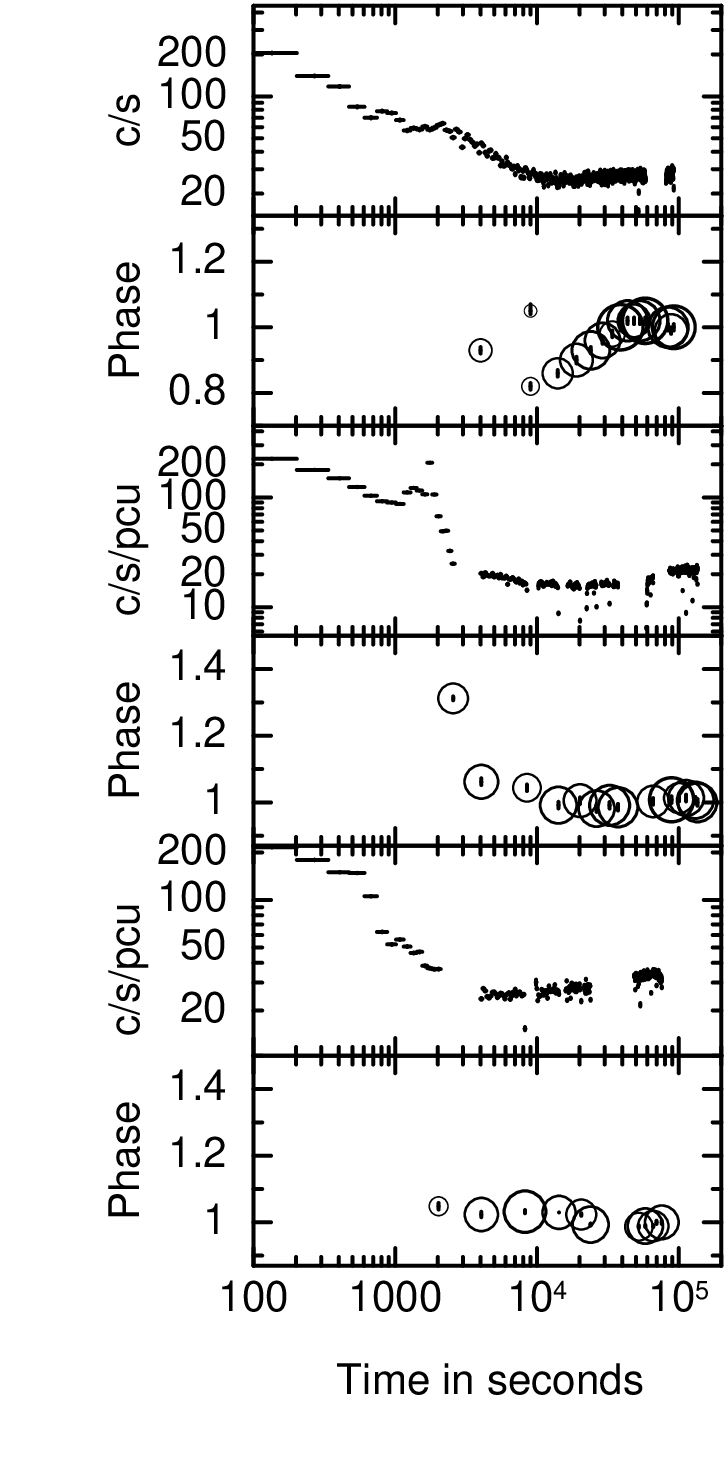}
 \caption{This figure illustrates the emergence and evolution of the dip phase and depth with time.
 Top panel: Plotted light curve binned with a binsize of 135~seconds.
 The bottom panel shows the evolution of dip phase and dip depth in the pulse profiles.
  To notice a change in phase depth with time each circle is scaled by a factor of 20.
  ~The x and y-axis is in log scale.
  For details see text.}
 \label{EPIC-phase-time}
 \end{figure*}

We divided the data into narrow time segments 
to separate the flares and the periods in between the flares, 
labelled as $1,2,3,4$ and so on (see Figure-\ref{LC}).
We have performed a pulse profile evolution study in the 2-10~keV band
because the dips appeared in the pulse profiles above 2~keV,
created using the persistent state data (refer Figure~\ref{pp}). \\
\par
 Figure~\ref{pp-evol} shows the evolution in the pulse profiles, created 
using data from the \textsc{EPIC}-pn.
The most important and interesting observations from Figure-\ref{pp-evol}~(\textsc{EPIC}-pn data) are described below: 
\begin{itemize}
 \item 
 Pulse profiles during the flares (segments $2$, $4$, $6$) and just before and after the flares (segments $1$, $3$ and $5$)
 are nearly sinusoidal, but the peak of the profiles was found to 
 have a significant phase offset~(of about $0.23\pm0.02$).
 In particular, we note that the time
 segments around the flares do not have the dip feature even though the mean photon count rates in segments $1$, $3$ and $5$ are similar to those of the persistent emission.
 \item Near the end of last flare and the beginning of persistent emission (segment~$8$),
  broad dip like features began to emerge.
 \item Data during the segment $9$ showed further development of the dip near phase $0.9$, though the dip
 is not very deep and narrow in comparison to the dip seen in the profiles created using later segment $10$ of the persistent emission.
 Moreover, some other dip-like features are also observed
 in the profiles created using data of segment $10$.
 \item Pulse profiles of segment $11$ (the region after the eclipse) exhibited many dip-like structures in it.
 The most prominent dips were observed near phases $0.2$~and~$1.0$ with the rest of the profile also showing some
 structures in it.
\end{itemize}
From the above-mentioned sequence, we infer that dips are formed in the pulse profile
after all the flares are completed, and it takes several thousand seconds for the dips
to settle in phase. 
Therefore, in order to measure the timescale required for the formation of the dip, we further divided segment
$9$ into narrow segments, each with 5~ks of exposure time (see right
plot of Figure-\ref{pp-evol}).
It is evident from the figure that for the first
five thousand seconds of segment $9$ there are several evolving structures in the profile
with the most prominent ones near phase $0.8$ and $1.0$.
    Profiles created using the next five thousand seconds of the segment $9$ show a
    sharp dip near phase 0.8. During later times of the $9$th segment the profiles show the appearance
    of a clear dip near phase 1.0. \\
    \par
For a quantitative estimation of the timescale of formation of a dip,~the
 location of dip phase and its depth, we created pulse profiles using the data
in short segments starting from the segment $8$.
Dips observed in the phase range of $0.8-1.1$ in each of these profiles were fit to a constant 
and a negative Gaussian.
~This robust method of finding the exact location of the dip phase and its depth
allowed us to study the dip evolution more accurately.
Figure-\ref{EPIC-phase-time} shows the evolution of dip phase and depth with time,
where the top panel shows the light curve starting 100~seconds after the peak of the last flare while
the bottom panel gives the measure of the dip phase and its depth 
as observed in the profiles at the corresponding times.
After the flares observed during this \emph{XMM-Newton} observation, it took
nearly 4000~seconds starting from the peak of the last flare for dip feature
to emerge.~However, it took more than 30,000 seconds for the dip to settle in phase.~The depth
of the dip which is represented by the size of circle in the lower panel of Figure-\ref{EPIC-phase-time}
indicates that for about 9000~seconds dips 
were quite shallow in comparison to the rest of the dips seen in the pulse profiles. \\
\par
It is interesting to notice that although the flare decay timescale
varies from flare to flare, the time difference between the consecutive flares
is approximately 4~ks. \\
\par

\begin{figure*}
\centering
\begin{minipage}{0.45\textwidth}																															
\includegraphics[height=5.5in,width=7.cm,keepaspectratio]{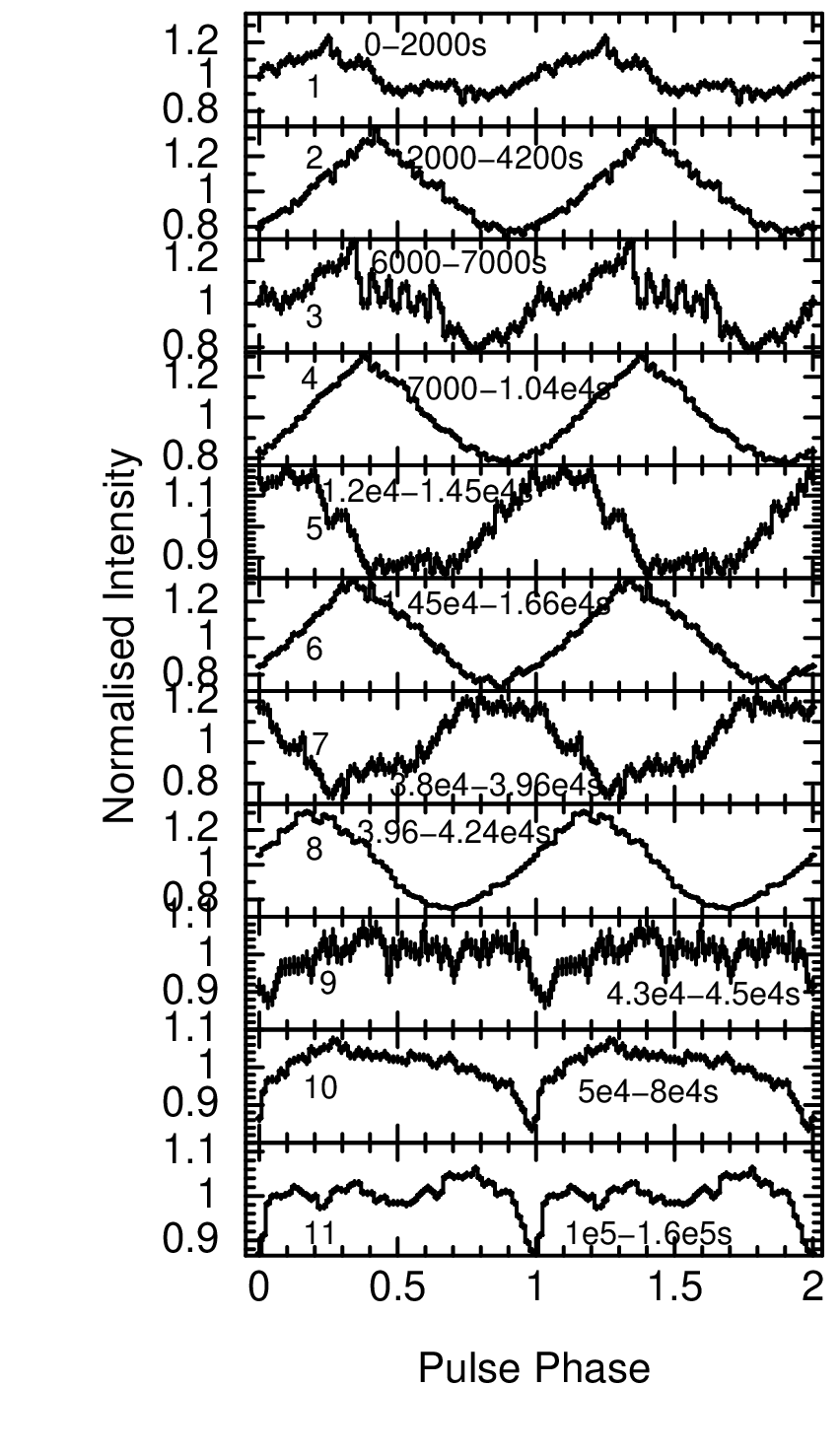}
\end{minipage}
\hspace{0.05\linewidth}
\begin{minipage}{0.45\textwidth}
\includegraphics[height=5.5in,width=7.cm,keepaspectratio]{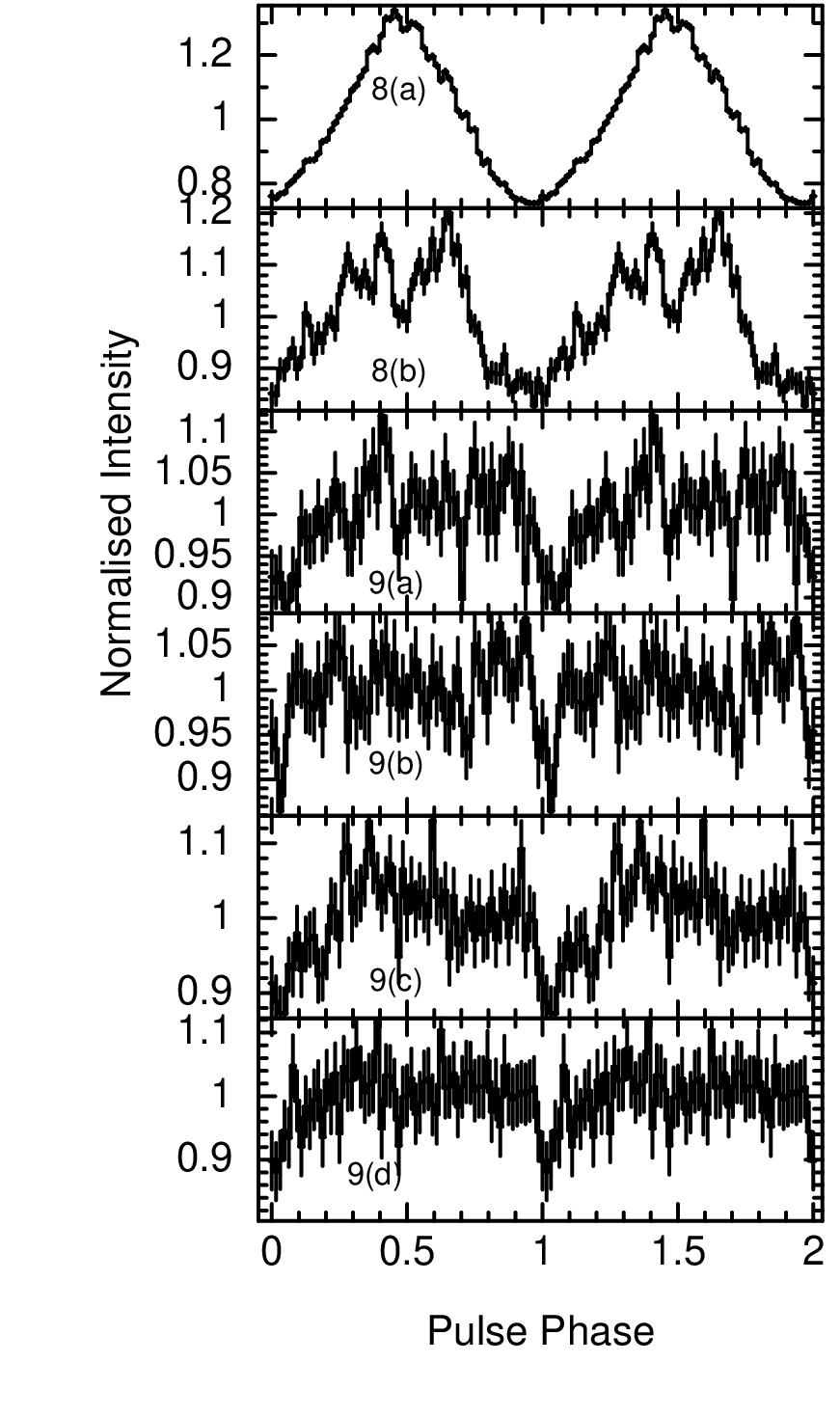} 
\end{minipage}
 \caption{The left-hand plot shows the pulse profiles created using \textsc{PCA} data (OBSID-P10135)
 in short segments as marked in middle panel of Figure-\ref{LC}.~Each panel is also marked with the time range for each segment.
These profiles are created using the time series in the 2-10~keV band and are binnned into 64 phasebins while the plot on the right 
shows the pulse profiles created using the sub-segments of segments $8$ and $9$.~~The duration of segment~8(a), 8(b) is 2000 and 800~seconds, respectively.
Segment 9 is
divided into 4 each having 500~seconds of exposure time.~Pulse profiles of the sub-segments
of segment $9$  are binned into 32 phasebins.}
\label{PP-RXTE}
\end{figure*}

\begin{figure*}
\centering
\begin{minipage}{0.45\textwidth}
\includegraphics[height=5.5in,width=6.5cm,keepaspectratio]{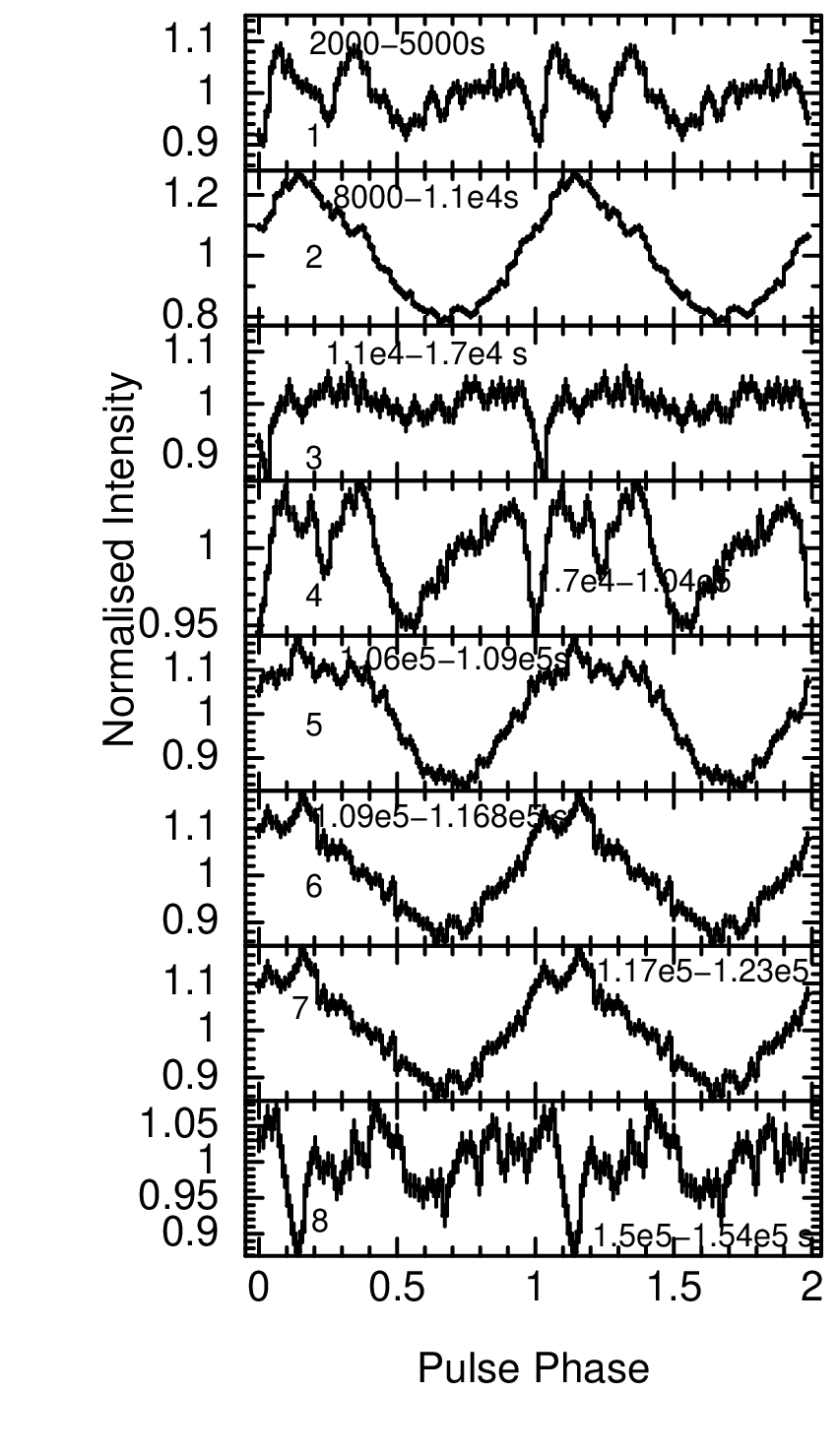}
\end{minipage}
\hspace{0.05\linewidth}
\begin{minipage}{0.45\textwidth}
\includegraphics[height=5.5in,width=6.5cm,keepaspectratio]{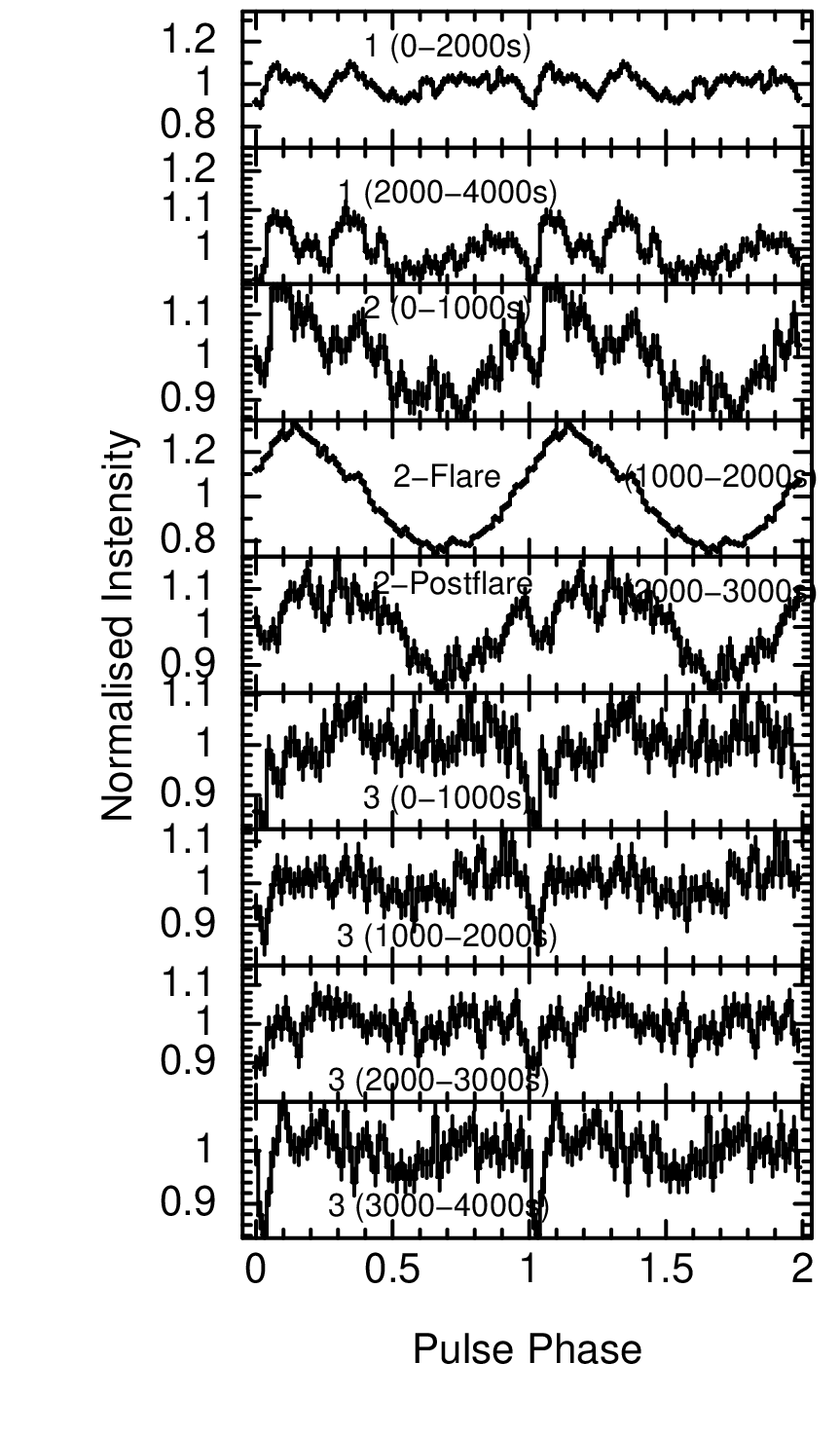} 
\end{minipage}
 \caption{Left hand side plot shows the pulse profiles created in short segments of \textsc{PCA} data with OBSID-P40064,
 as marked in the bottom panel of Figure-\ref{LC}.~Each panel is also marked with the time range used.
These profiles are created using the time series in the 2-10~keV band and are binnned into 64 phasebins.
The plot on the right side
shows the pulse profiles created using the sub-segments of segments $1$, $2$ and $3$.~The pulse profiles of the
sub-segments of segment $3$ are binned into 32 phasebins.}
\label{PP-RXTE-40064}
\end{figure*}

Pulse profiles created with short segments of the other two \textsc{PCA} observations~-~P10135 
and P40064 are shown in a left hand plot of Figures~\ref{PP-RXTE} and \ref{PP-RXTE-40064}
respectively.~In both these observations 
profiles during the flares are simple and exhibit sinusoidal shapes.
However, it is interesting to notice that the low intensity flares~(segments $6$ and $7$)
seen in the observation with ID-P40064 exhibits a signature of a shallow dip near the peak
of the profiles (Figure~\ref{PP-RXTE-40064}).
The phase offset of $0.21\pm0.01$ is also observed between the peak of the
pulse profiles of segments $1$, $3$, $5$ and $7$~(just before and after the flares)
and the flare pulse profiles (Figure~\ref{PP-RXTE}). \\
\par

%
%
%

The plot on the right hand side of Figures~\ref{PP-RXTE} and \ref{PP-RXTE-40064} shows a detailed
investigation of the evolution of the dip in the pulse profiles after the end of flares.
From Figure~\ref{PP-RXTE}, we observe that near the decline of the last flare $8(b)$ of P10135, the profiles showed some emerging
features in comparison to the flare profile~($8a$).
The pulse profiles created using the subsegments of the segment $9$ (each with 1200~seconds time)
also showed the evolution of a dip near phase 1.0 with time.
This result is consistent with that obtained 
using the \textsc{EPIC}-pn data. 
For the observation-P40064, 
we again noticed that the dip near phase 1.0 is evolving with time and this feature 
became deep and narrow in the last subsegment of $3$. Since flares are not continuous
for this particular \textsc{PCA} observation, we could not measure the exact time difference between 
the two consecutive high intensity flares.~Hence, structures seen in the profiles near the beginning of the flare $2$
could not be explained with this observation.  \\
\par
For a quantitative measurement of the dip phase and depth, we followed the same technique~(as explained for \textsc{EPIC}-pn data)
of fitting a dip~(in the phase range of 0.9-1.35 for P10135 and 0.95-1.1 for P40064) with a constant and a negative gaussian model
component (Figure~\ref{EPIC-phase-time}).
From the Figure~\ref{EPIC-phase-time}, it was found that it took nearly 2500~seconds after the peak of
a last flare for the dip to emerge.~The time difference between 
two the flares is also nearly 2500~seconds.~This is again consistent with the fact that no dip is seen in the persisent
emission~($3$,$5$) between two flares.~The dip-phase settle after about 20,000 seconds.\\
However, for the observation-P40064, it was found that dip occured about 2000~seconds
after the peak of first very high intensity flare~(segment $2$).
Both the \textsc{PCA} observations also showed that it takes a few thousand of seconds
for dip to develop after the end of the flares. \\

\begin{figure*}
\centering
\begin{minipage}{0.45\textwidth}
\includegraphics[height=6.5in,width=6.5cm,keepaspectratio]{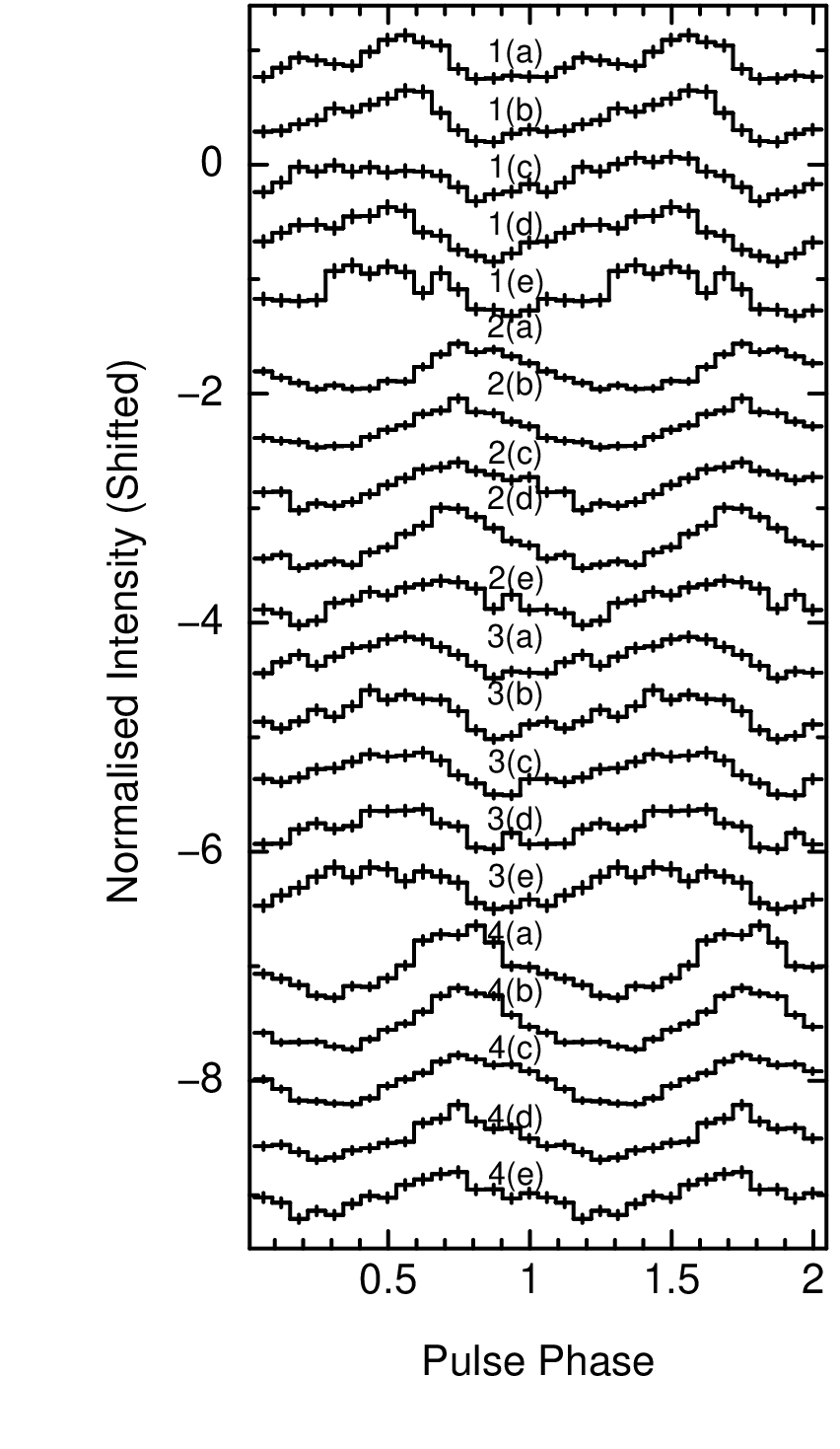}
\end{minipage}
\hspace{0.05\linewidth}
\begin{minipage}{0.45\textwidth}
 \includegraphics[height=6.5in,width=6.5cm,keepaspectratio]{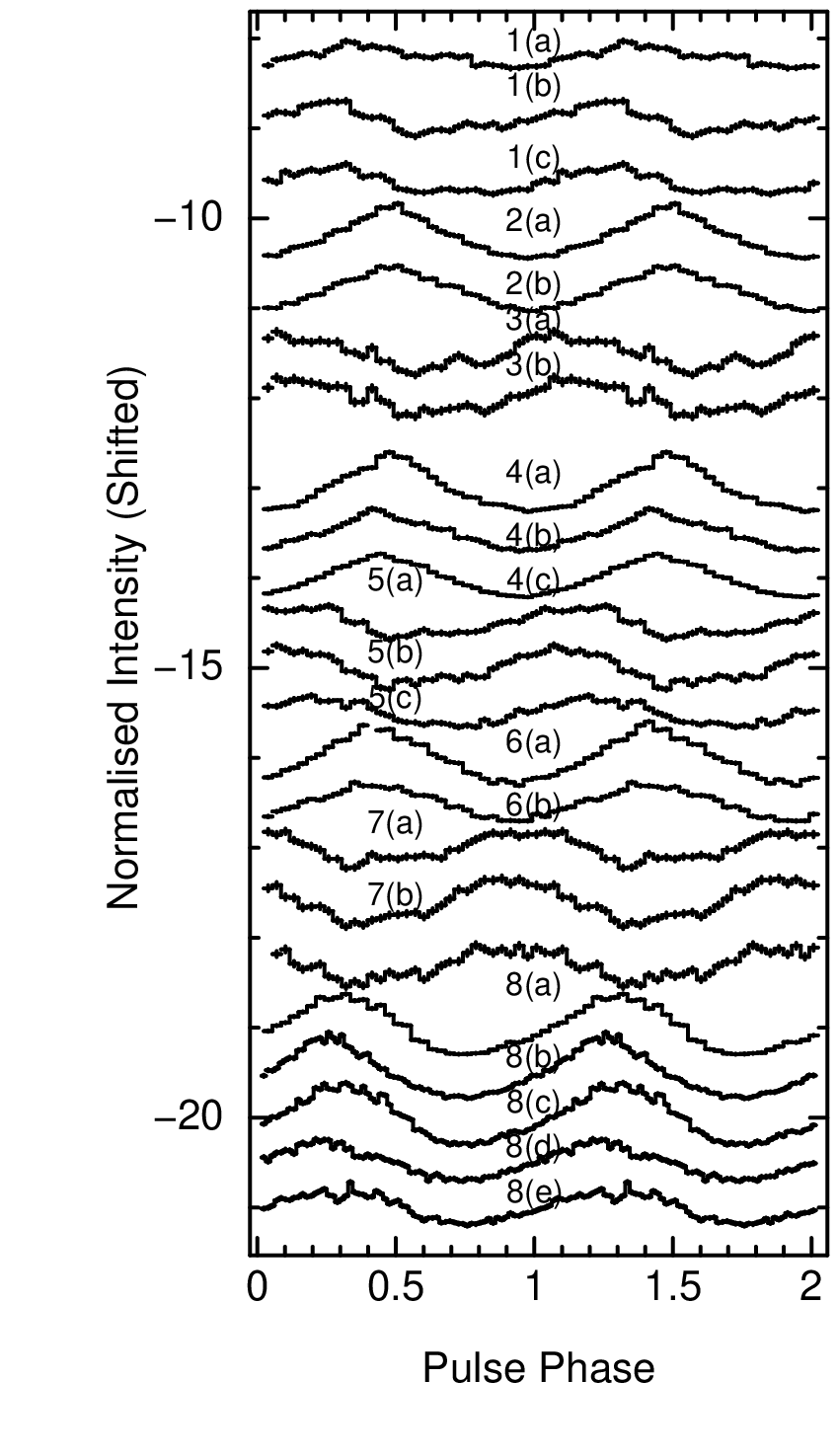}
\end{minipage}
\caption{Left hand plot shows pulse profiles created using sub-segments of $1$, $2$, $3$ and $4$ intervals of the
top panel of Figure-\ref{LC}~(\textsc{PN} data), while the right-hand plot shows the pulse profiles created using sub-segments
of $1$, $2$, $3$, $4$ $5$, $6$, $7$ and $8$ 
shown in the middle panel of Figure-\ref{LC}~(\textsc{PCA}-P10135), 
each vertically shifted for ease of viewing. The profiles are binnned into 16 phasebins.}
\label{pp-ABCD}
\end{figure*}

\subsection{Phase Shift During Flare Profiles }
To keep track of the phase shift observed between the peak of the pulse profiles
during flares~(segments $2$, $4$ and $6$) and just before and after flare segments (segments $1$, $3$~and $5$)
in the \textsc{EPIC}-pn data,
we further divided $1$, $2$, $3$ and $4$ segments in the short intervals, shown in the left panel of Figure-\ref{pp-ABCD}. 
It was found that the pulses are in phase during intervals $1(a)-1(e)$.
Then there is a sudden phase change at the beginning of the
flare at interval $2(a)$, the pulse peak appears delayed (i.e., the beaming
during the flare is away from the direction of rotation).
The phase seems to be constant during all the segments of
flare~($2$) and returns by a sudden jump to the original phase at $3(a)$.
$3(a)-3(e)$ have a similar phase, just like in $1(a)-1(e)$.
The same kind of sudden phase shift happens at $4(a)$.
The phase shift observed between the flares and pre-flares measured using the location 
of the peaks in Figure~\ref{pp-ABCD} varies between 0.2 and 0.5.
\\

For the \textsc{PCA} observation P10135, 
we divided segments $1$, $2$, $3$, $4$, $5$, $6$, $7$, $8$ 
into short intervals. Pulse profiles created using the sub-segments are shown in right hand side plot of Figure-\ref{pp-ABCD}.
The same range of phase shift between the peak of pulse profiles is also observed in
this observation.
This confirms the change in beaming pattern within flares.

\section{Spectroscopy with \textsc{EPIC}-pn}

\subsection{Phase-Averaged Spectroscopy}
Pulse-phase averaged spectral analysis was performed using the mean spectrum extracted using the persistent emission data 
(excluding flares and an eclipse). 
The spectrum was rebinned by factor of 4 between 1-10~keV 
using ftool \emph{grppha}. The spectral fitting in the 0.3-10~keV band was
done using \emph{Xspec Version}: 12.8.0 \citep{Arnaud96}.  
We have first tried to fit the continuum of the phase-averaged spectrum using as model components:
bremmstrahlung for the soft excess and a powerlaw, each attenuated with line of sight
absorption. The minimum of the interstellar absorption component was set at the value of 
the Galactic column density towards the source.
A bremmstrahlung model component alone did not provide an acceptable fit, some soft excess was still seen
in the residual.
Therefore, we added an additional blackbody emission component \citep{Woo96} to fit the low energy excess. \\
\par
\begin{figure*}
\includegraphics[height=5.5in, width=5.5in, angle=0,keepaspectratio]{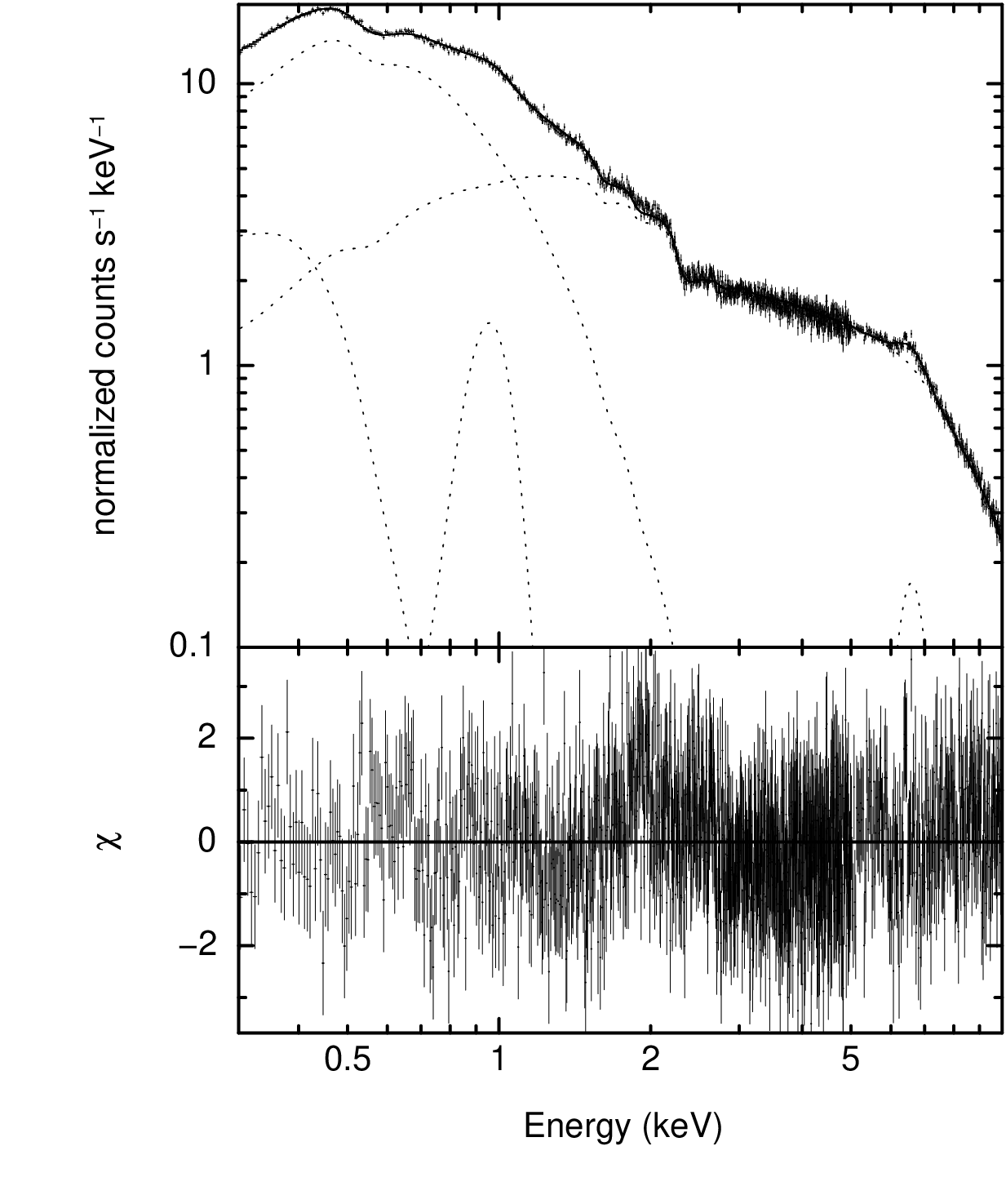}
 \caption{Data and folded model spectrum of LMC~X--4 from \emph{XMM-Newton}~\textsc{EPIC-PN}.}
\label{Average}
\end{figure*}

Using only the continuum model still showed an excess in the form of an emission line features around 1.0~keV and 6.4~keV.
The \emph{RGS} data from the same observation was used by \citet{Neilsen09}, 
where they have reported a broad emission feature near 1~keV. Therefore, considering this fact we added a gaussian 
component centered around 1~keV with a width of $0.080$ \citep{Neilsen09}.
An additional gaussian feature was also added with the line energy
centered around 6.4~keV (see Figure~\ref{Average}). 
The parameters obtained from the fitting are given in Table-2. These parameters
are in good agreement with the values reported by \citet{Neilsen09} using the \emph{RGS} data from the same observation
(see Table-\ref{Neilsen}). These authors also fixed the galactic neutral hydrogen column density at 5.78 $\times 10^{20} \it{cm}^{-2}$.

\begin{table}
\centering
      	 \caption{Best-fitting parameters of Phase-averaged spectrum of LMC~X--4.}
         \begin{tabular}{ l  l }
         \hline
         \hline
           
         Parameter & Model Values  \\ 
         N$_H$ (10${^2}{^2}$atoms cm$^{-2}$) & $0.06$  \\
    	 
         $\Gamma$  & $0.750\pm{0.005}$ \\ 
         
        $N_{PL}^a$  & $(5.65\pm{0.04})\times10^{-3}$  \\

         $kT_{bbodyrad}~(keV)$  & $0.043\pm{0.001}$ \\
        
         $kT_{Bremss}~(keV)$  & 0.400$\pm{0.004}$  \\
         $\rm{Fe_{E}^b}$   & $6.62\pm0.02$ \\
         $\rm{Fe_{W}^c}$   & $0.40 \pm 0.02 $ \\
         $\rm{Fe_{flux}^d}$  & $2.3\pm{0.1}$ \\
          $\rm{Fe_{EW}}^e$   & $0.17\pm{0.01}$ \\
           Reduced $\chi^2$  &   1.2~(dof~1179)   \\
           
  \hline
  \end{tabular}

 \bigskip

{\bf{Notes}}: Errors quoted are for the 68 $\%$ confidence range. \\
       \hspace{0.5in} a $\rightarrow$ Powerlaw normalisation~($N_{PL}$)
       is in units of $\rm{photons~cm^{-2}~s^{-1}~keV^{-1}}$ at 1~keV \\
       b $\rightarrow$ Iron-line energy in units of keV. \\
       c $\rightarrow$ Iron-line width in units of keV.  \\
  \hspace{0.2in}     d $\rightarrow$ Gaussian normalisation is in units of $10^{-4}~\rm{photons~cm^{-2}~s^{-1}}$ \\
   \hspace{0.5in}    e $\rightarrow$  Iron-line equivalent width in units of keV. \\
            \end{table}

    \begin{table}
\centering
      	 \caption{Best-fitting parameters obtained by \citet{Neilsen09}}
         \begin{tabular}{ l  l }
         \hline
         \hline
           
         Parameter & Model Values  \\

         $\Gamma$  & $0.813\pm{0.007}$ \\

         $kT_{bbodyrad}~(keV)$  & $0.043^{+0.001}_{-0.0002}$ \\
        
         $kT_{Bremss}~(keV)$  & $0.455^{+0.022}_{-0.001}$  \\

  \hline
  \end{tabular}

 \bigskip

\label{Neilsen}
  
            \end{table}

\begin{figure*}
\centering
\begin{minipage}{0.45\textwidth}
\includegraphics[height=5.75in,width=6.5cm,keepaspectratio]{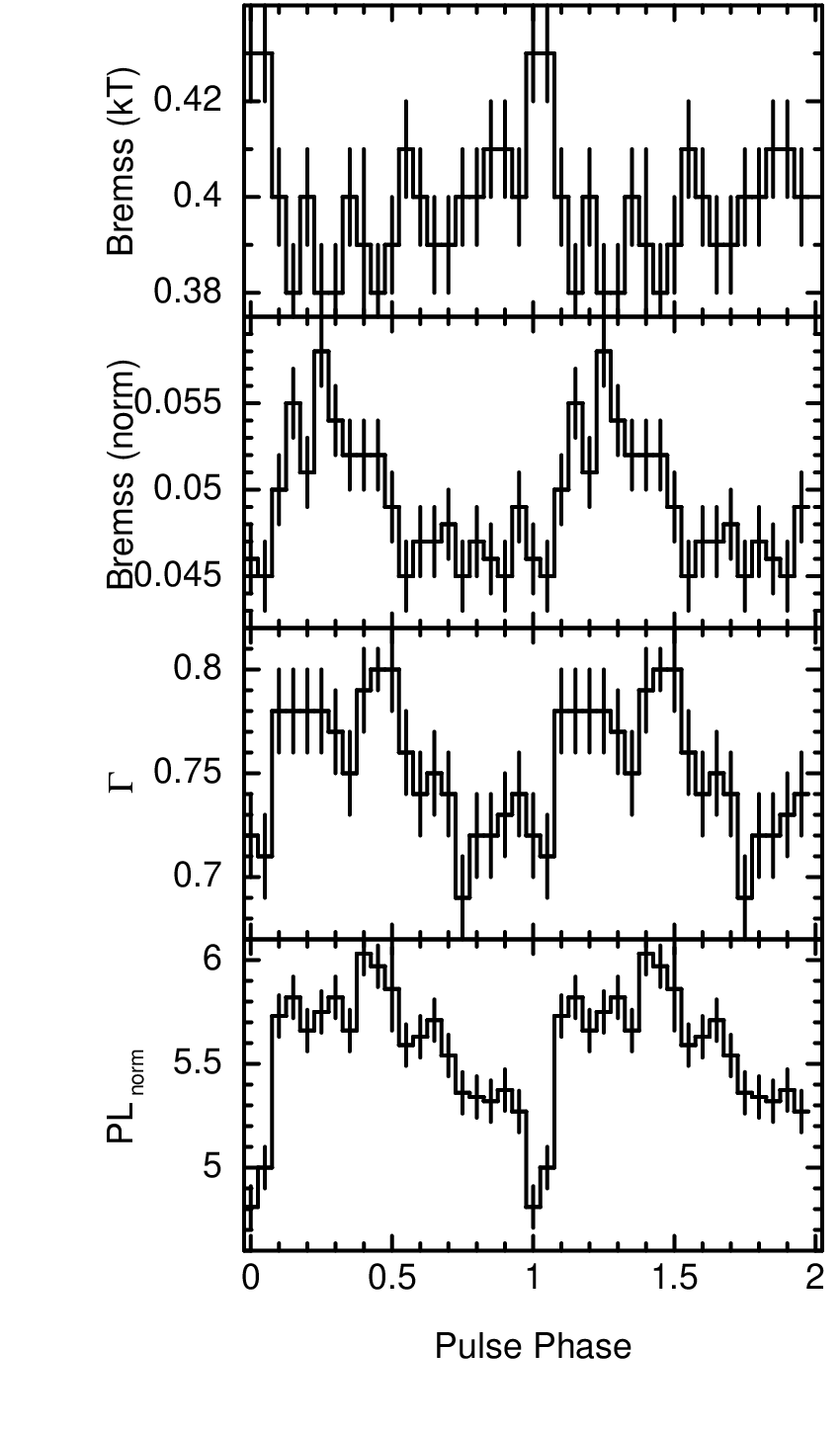}
\end{minipage}
\hspace{0.05\linewidth}
\begin{minipage}{0.45\textwidth}
\includegraphics[height=5.6in,width=6.5cm,keepaspectratio]{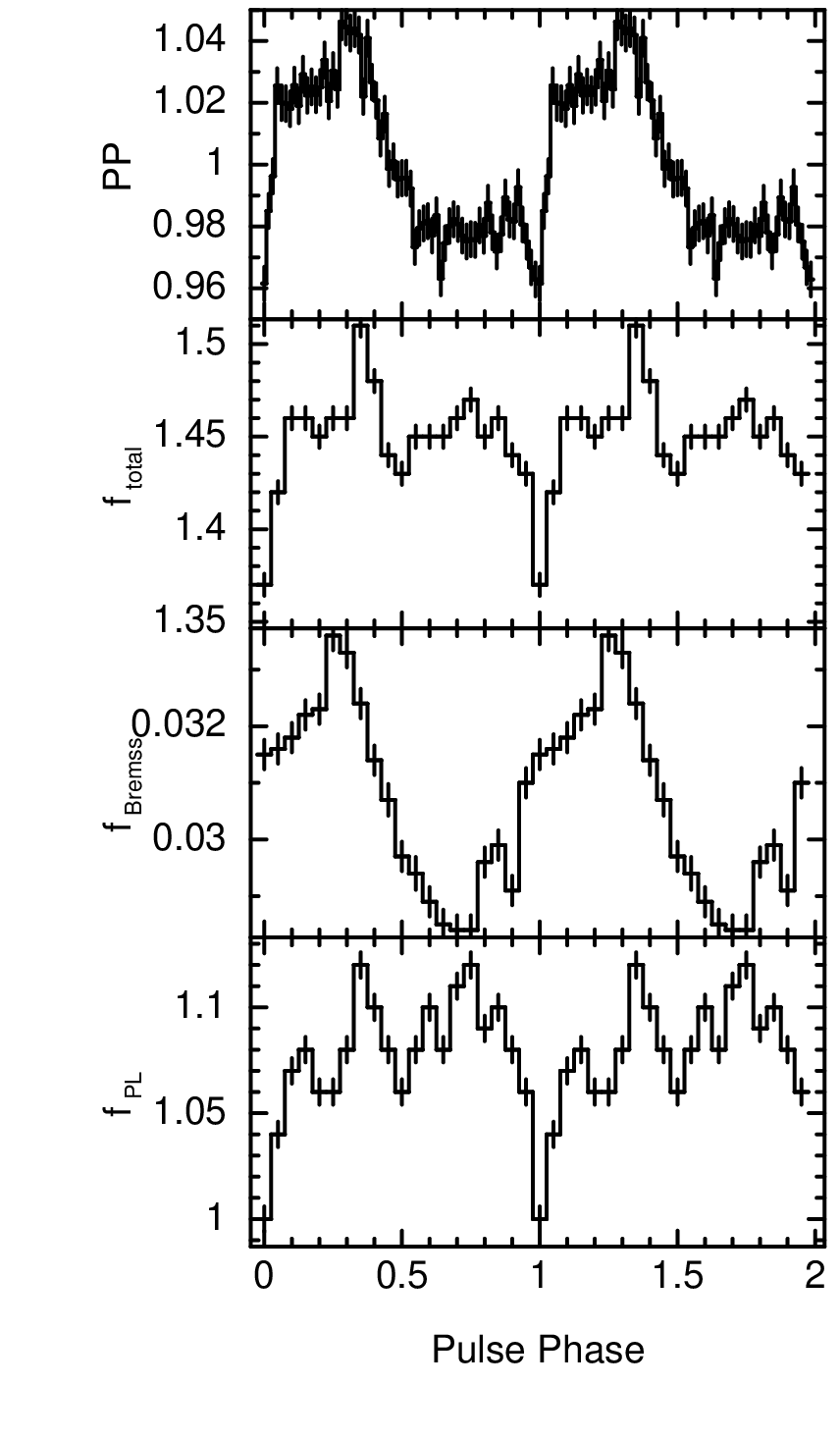}
\end{minipage}
\caption{Left:~Variations of the continuum spectral parameters accross the pulse phase.
 Powerlaw normalisation~($PL_{norm}$)
       is in units of $\rm{photons~cm^{-2}~s^{-1}~keV^{-1}}$ at 1~keV. 
Right:~variation of the flux~(f) of the continuum.
$f_{total}$, $f_{Bremss}$ $\&$ $f_{PL}$ are measured in units of $10^{-10}$$ergs~cm^{-2}~sec^{-1}$.~
All the errors were estimated with 1$\sigma$ confidence.}
\label{soft}
\end{figure*}

\subsection{Phase-Resolved Spectroscopy}
The pulsating nature of the soft spectral component during its high state has been studied by \citet{Woo96,Paul02,Paul04}
using data of \emph{ROSAT}, \emph{Ginga}, \emph{ASCA} and \emph{BeppoSAX}.
A significant difference in the pulse profile below and above
2~keV of the \textsc{PN} light curve (Figure-\ref{pp}) also indicates
the soft component of spectrum to pulsate differently compared to
the powerlaw component.
We here present phase resolved studies performed at narrow phasebins.

The event file used for the extraction of phase resolved spectra was corrected
for the binary motion as mentioned earlier in section-2.
Phase resolved spectra were created using the 'phase' filter of \emph{ftools} \footnote{http://heasarc.gsfc.nasa.gov/ftools/}
task \textsc{xselect} with a phase bin of 0.05.
We have used the same background spectrum and response files as were used for the phase-averaged spectrum.
X-ray spectra in
different phases were grouped using \emph{ftools} task \textsc{grppha} with minimum 25 counts per bin. For fitting the phase-resolved spectra
we followed exactly the same technique as we opted for the phase-averaged spectrum. 
However, we fixed all the line energies and line widths
to the values obtained in the best fit of phase-averaged spectrum. 
It is interesting to observe the pulsations (pulse fraction $17\%$)
of the soft component (thermal bremmstrahlung) in narrow phase bins
(see right plot of Figure-\ref{soft}). The profile shapes were similar to those reported by \citet{Paul04}.
The powerlaw flux also exhibited variation accross the pulse phase with some similarity with the pulse profiles
in the 5-10~keV band. A dissimilar pulse profile between the low and high-energy
bands is similar to the earlier results obtained from \emph{ASCA} and \emph{BeppoSAX}
 that indicated that the
soft component may have a different origin of emission, or that the geometry of
emission is different in different energy ranges \citep{Paul02,Paul04}.
The continuum parameters like the bremmstrahlung temperature, its normalisation and powerlaw index also
exhibit some variation accross the pulse phases (Figure-\ref{soft}). 
 
\section{Summary and Discussions}
In this work,
we have probed the pulse profiles of LMC~X--4, during the flares
and outside the flares.  \\
A study performed using the data from \emph{XMM-Newton}
and \emph{RXTE} showed the existence of dips during persistent emission,
after the flares.
The narrow dips, observed in the pulse profiles, are believed to occur due to phase-locked
absorption of the emitted radiation by the optically thick material in 
the accretion column \citep{Cemeljic98, Galloway01, Naik11, Maitra_13, Devasia14}.
Therefore, the dipping structures seen in the pulse profiles
during persistent emission indicate the existence of an absorption stream in the accretion column
that causes the drop in intensity.
The observed broad pulse profiles during flares support
their origin due to gravitational bending of the flare emission which is
transported via fan beams.
Thus, the observed differences in the pulse profiles during and after the flares suggest
that significant change in the accretion stream happens during the transition between
flares and the persistent state.
The existence of the observations containing both flares and persistent emission
allowed us to estimate the timescales required for the formation of 
accretion stream that causes dips, after the accretion region and beaming etc
are disturbed during flares.
If flares occur due to sudden infall of additional matter then the peak of a flare should indicate
the time when this sudden inflow is coming to an end. Hence,~the time since the last 
flare peak would indicate the time since it started getting back to persistent emission.
From the analysis with \textsc{EPIC}-pn~(Figure-\ref{EPIC-phase-time}), we found that 
the time required for
the formation of accretion stream that causes the dip is nearly 4~ks starting from the peak of 
last flare.~Moreover, this observed time is nearly equal to the time difference between the
two flares.
The time estimated from the \textsc{PCA}
observation wih ID-P10135 is nearly 2500~seconds, this time is also
consistent with time difference between two flares.
The second \textsc{PCA} observation used in this work with ID~P40064
also indicated that the time required for the formation of a dip 
after the peak of flare is approximately 2000~seconds.
Hence, we conclude that the formation of a dip in the pulse profiles
is not abrupt, it takes a few thousand seconds for the formation
of a accretion stream that causes the dip. 
To our knowledge, this is for the first time such a measurement has been made for any pulsar.
The timescale of formation of the dips is significantly larger than the dynamical timescale of Keplerian motion
of the inner accretion disk and free fall timescale from the inner disk. \\
\par
Another very interesting result we have found in this study is the existence of a significant
 phase shift between the pulse profiles
from the persistent emission (just before and after the flares) and the flares,
with the \textsc{EPIC}-pn data.
This suggests an evolution of the beaming pattern as the flare evolves and decays. 
Similar behaviour was also observed in one of the \textsc{PCA} observations~P10135.\\
\par
Pulse profiles during flares in LMC~X--4 are known to exhibit simple 
shapes \citep{Levine91, Levine00, Moon03} and they are known to
exhibit the same shapes in different energy bands \citep{Levine00}.
However, the profiles created using the \textsc{PCA} observation with ID~P40064
provide evidence in support of intensity dependence of the flare pulse profiles. 
It was observed that low intensity flares~(e.g., $6$ and $7$ segment)
show the presence of a shallow dip near the peak of the profiles.
However, this structure was not observed in high intensity flare pulse profiles. \\
\par
Very complex changes in the pulse profiles with luminosity have also been observed in pulsars like
EXO~2030+375 \citep{Parmar89}, GX~1+4 \citep{Paul97}, Cepheus X-4 \citep{Mukerjee00}.
\citet{Parmar89} suggested one possible explanation for the luminosity dependence
of profiles to be switching of the beaming mechanism from fan to pencil configurations and vice-versa.
The same hypothesis was applicable for the change in the pulse shape observed for Cep X-4
during the \textsc{IXAE} observation \citep{Mukerjee00}.
As mentioned earlier, we observe simple sinusoidal profiles during flares 
and complex features like dips in the pulse profiles, using the data during persistent emission.
Therefore, we think that the changes in the pulse profile during and after the flares
indicate changes in the shape and beaming pattern of the accretion column / emission region.
This behavior is, however, different from that observed in SMC~X--1.~\citet{Moon03a} found simple and
broad profiles during both the flares and non-flaring states of SMC~X--1.
The difference between SMC~X--1 and LMC~X--4 could be related to the difference in the strength of the magnetic field. \\

Pulse phase resolved spectroscopy carried out in narrow phasebins 
confirmed the presence of pulsations in the soft spectral
component. The pulse profile shape is sinusoidal for the soft component and
is different from that seen in the hard component. 
The pulsating nature of the soft component, though it was measured earlier 
with \emph{ASCA}, \emph{BeppoSAX}, is more clear in this \emph{XMM-Newton}
observation due to its higher sensitivity.~This indicates that the soft and hard
emissions have different origins.

\section*{Acknowledgments}
The research has made use of data obtained from High Energy Astrophysics
Science Archive Research Center (\emph{HEASARC}). 
A.B is grateful to the Royal Society and SERB~(Science $\&$ Engineering Research Board, India)
for financial support through Newton-Bhabha Fund.

%


\newpage
\bibliographystyle{apj}

\bibliography{complete-manuscript}

\end{document}